\documentclass[aps,amsmath,onecolumn,superscriptaddress,prl]{revtex4}

\usepackage{amsmath,amssymb,bm}
\usepackage[dvips]{graphicx,color}
\usepackage{braket}
\usepackage{ulem}
\usepackage{setspace}[10]
\usepackage{comment}
\usepackage[dvips]{color}
\usepackage[dvipsnames]{xcolor}
\usepackage{comment}
\usepackage{mathrsfs}

\newcommand{\bfig}{\begin{figure}\begin{center}}
\newcommand{\efig}{\end{center}\end{figure}}

\begin{document}
\title{Spin-current version of solar cells in non-centrosymmetric magnetic insulators}
\author{Masahiro Sato}
%\email{masahiro.sato.phys@vc.ibaraki.ac.jp}
\affiliation{Department of Physics, Ibaraki University, Mito, Ibaraki 310-8512, Japan}
\author{Hiroaki Ishizuka}
\affiliation{Department of Applied Physics, The University of Tokyo, Bunkyo, Tokyo 113-8656, Japan}
\date{\today}
\begin{abstract}
Photovoltaic effect, e.g., solar cells, converts light into DC electric current. 
This phenomenon takes place in various setups such as in noncentrosymmetric crystals and semiconductor pn junctions.
Recently, we proposed a theory for producing DC spin current in magnets using electromagnetic waves, 
i.e., the spin-current counterpart of the solar cells. Our calculation shows that the nonlinear conductivity 
for the spin current is nonzero in a variety of noncentrosymmetric magnets, 
implying that the phenomenon is ubiquitous in inversion-asymmetric materials with magnetic excitations. 
Intuitively, this phenomenon is a bulk photovoltaic effect of magnetic excitations, 
where electrons and holes, visible light, and inversion-asymmetric semiconductors 
are replaced with magnons or spinons, THz or GHz waves, and asymmetric magnetic insulators, respectively. 
We also show that the photon-driven spin current is shift current type, and as a result, the current is stable 
against impurity scattering. 
This bulk photovoltaic spin current is in sharp contrast to that of well-known spin pumping that takes place 
at the interface between a magnet and a metal.
\end{abstract}
\maketitle

%%%%%%%%%%%%%%%%%%%%%%%%%%%%%%%%%%%%%%%%%%%%%%%%%%%%%%%%%%%%%%%%%
%%%%%%%%%%%%%%%%%%%%%%%%%%%%%%%%%%%%%%%%%%%%%%%%%%%%%%%%%%%%%%%%%
%%%%%%%%%%%%%%%%%%%%%%%%%%%%%%%%%%%%%%%%%%%%%%%%%%%%%%%%%%%%%%%%%
\section{Introduction}
\label{sec:intro}  % \label{} allows reference to this section
The technology based on electromagnetic waves and laser has long contributed to 
the development and discovery of photo-induced phenomena. 
Among them, the DC photocurrent in solar cells is a representative phenomenon~\cite{Sturman92}. 
The basic structure of solar cells is a junction of two different (p- and n-type) semiconductors.  
The point is that the inversion symmetry breaking is necessary to create a DC current by AC electromagnetic fields 
(see Appendix). 
Electromagnetic wave is merely an oscillating field and thereby it does not favor any spatial direction. Thus, the system (i.e., solar cell) should have a peculiar direction to induce electric current. Besides the pn junctions, single bulk crystals without inversion symmetry (i.e., noncentrosymmetric crystals), such as perovskite photovoltaics, 
have attracted attention as another platform of photovoltaic devices. 
The photocurrents in semiconductors~\cite{Kohli11,Cote02,Sotome19}, transition-metal oxides~\cite{Koch76,Dalba95} and organic materials~\cite{Nakamura2017} have been experimentally observed, and the corresponding microscopic theories~\cite{Kraut79,Belinicher82,Sturman92,Kral00,Sipe00} for photocurrents 
have been developed, especially, in recent years~\cite{Young12,Morimoto16,Cook17,Ishizuka17}.   
These phenomena are a consequence of the asymmetric optical transition of electrons. 

Besides electrons, photo-induced dynamics has also been studied in various systems with different quasi particles 
such as ferro and antiferromagnets~\cite{Kirilyuk10,Nemec18}. 
However, the optical control of the flow of these quasi particles has not been studied so far.  
In this work, we explored one such example in magnetic insulators, i.e., 
an optical generation of magnon or spinon flow in magnetic materials~\cite{Ishizuka19-1,Ishizuka19-2}. 
Since magnons and spinons are magnetic excitations carrying angular momentum, 
the photovoltaic magnon/spinon current is the spin-current version of photocurrent. 
 
Spin current~\cite{Maekawa12} is defined as the flow of angular momentum and 
have been continuously a hot topic in spintronics. 
It can be viewed as a new instrument of information processing besides usual electric current, 
and several methods of creating/controlling spin current have been proposed and realized.  
Moreover, spin current has gathered attention as an important element of novel transport phenomena 
from the viewpoint of fundamental science.

Intense terahertz (THz) or gigahertz (GHz) waves are efficient to create DC photo spin current 
in magnetic insulators because energies of magnetic excitations with spin angular momentum 
(i.e., carriers of spin current) are usually located in THz or GHz range. 
The THz-laser science has massively progressed in recent years, 
and now we can utilize strong THz laser pulse~\cite{Hirori11,Sato13,Liu17} whose AC-electric-field strength is 
beyond 1 MV/cm ($\sim$ 0.1-1T of AC magnetic field) in the range of 0.1-10 THz. 
Actually, several THz-laser (wave) driven magnetic phenomena have been actively studied. 
For instance, THz-wave driven magnetic resonance~\cite{Mukai2016,Baierl16,Kubacka14}, 
a magnetic-excitation mediated high harmonic generation~\cite{Lu17} with THz laser, etc. 
have been observed. Microscopic theories~\cite{Takayoshi14-1,Takayoshi14-2,Sato16,Fujita17,Ikeda19,Sato20} 
for THz-wave driven magnetic phenomena have also been developed recently.  
Similar technologies using THz waves apply to the photovoltaic spin current as well. 

This proceedings discusses the essential aspects of photo-generation of spin current in magnetic insulators. 
As shown in Table~\ref{tab:setup}, noncentrosymmetric semiconductors, electrons (or holes), 
and visible or infrared waves are replaced with 
noncentrosymmetirc magnets, magnetic excitations (magnons, spinons etc.), and GHz/THz waves 
in our proposal~\cite{Ishizuka19-1,Ishizuka19-2}.

\begin{table}[ht]
\caption{Set ups of solar cell of creating electric current and our proposal of doing spin current.} 
\label{tab:setup}
\begin{center}       
\begin{tabular}{|l|l|l|} 
\hline
\rule[-1ex]{0pt}{3.5ex}  Solar cell & Our proposal (spin-current version of solar cell) \\
\hline
\hline
\rule[-1ex]{0pt}{3.5ex}  Noncentrosymmetric semiconductors & Noncentrosymmetric magnetic insulators  \\
\hline
\rule[-1ex]{0pt}{3.5ex}  Infrared or visible light (1THz-1PHz) & GHz or THz waves (0.1-10THz)  \\
\hline
\rule[-1ex]{0pt}{3.5ex}  Electrons and holes  &  Magnetic excitations (magnons, spinons, etc.) \\
\hline
\end{tabular}
\end{center}
\end{table}

%%%%%%%%%%%%%%%%%%%%%%%%%%%%
%%%%%%%%%%%%%%%%%%%%%%%%%%%%
%%%%%%%%%%%%%%%%%%%%%%%%%%%%
\section{Modeling}
\label{sec:model}
In this section, we define our models for photo-induced spin current. 
As we discussed in the previous section, inversion-asymmetric systems are necessary to create the photo spin current.  
To this end, we consider two kinds of quantum spin models as simple, but realistic magnetic systems.

%%%%%%%%%%%%%%%%%%%%%%%%%%%%
%%%%%%%%%%%%%%%%%%%%%%%%%%%%
\subsection{Quantum Spin Chains}
\label{sec:chain}
\begin{figure} [ht]
\begin{center}
\begin{tabular}{c} %% tabular useful for creating an array of images 
\includegraphics[height=8cm]{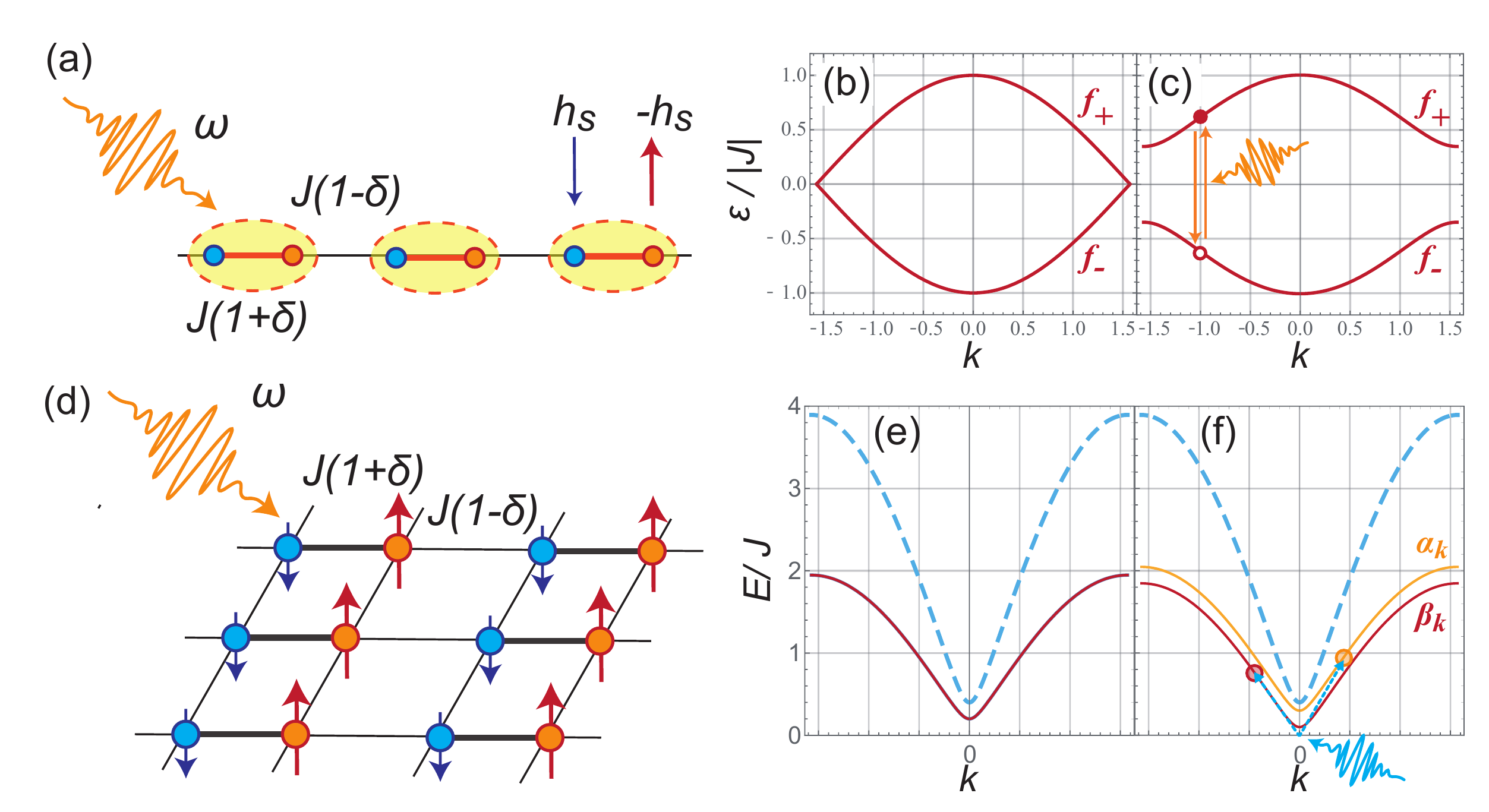}%{mcr3b.eps}
\end{tabular}
\end{center}
\caption{\label{fig:models} (a) Noncentrosymmetric quantum spin chain model and 
(d) noncentrosymmetric antiferromagnetic (or ferrimagnetic) insulator model. 
In the spin chain (a), a dimerization $\delta$ and a staggered field $h_s$ break the site- and 
bond-centered inversion symmetries, respectively. In the antiferromagnet (d), 
a dimerization $\delta$ and the ordering structure respectively break the site- and 
bond-centered inversion symmetries. Panels (b) and (c) represent 
the band dispersion of the JW fermion (here we call it spinon) for inversion-symmetric and asymmetric spin chains, 
respectively. Panels (e) and (f) depict the $\hat x$-direction magnon bands of the antiferromagnetic insulator models 
with $h_+=0$ and $h_+\neq 0$, respectively. For simplicity, we take the 1D limit $J_\perp/J\to 0$. 
The dotted blue curve correspond to 
$E_\alpha(\textit{\textbf{k}})+E_\beta(\textit{\textbf{k}})$.}
\end{figure}

First, we consider an inversion-asymmetric quantum spin chain~\cite{Giamarchi03,Gogolin04,Tsvelik04} 
whose Hamiltonian is given by
\begin{align}
  H_{\rm 1D}=&\sum_j J(1+(-1)^j\delta)(S_j^xS_{j+1}^x+S_j^yS_{j+1}^y) -\sum_j (h+(-1)^jh_s) S_j^z, 
\label{eq:chain}
\end{align}
where $S_j^{x,y,z}$ are $S=\frac{1}2$ spin operators on $j$th site ($\hbar$ is set to be unity), 
$J$ is the easy-plane exchange coupling whose energy scale is usually in GHz or THz regime, 
and $h$ is the external, uniform magnetic field along the $z$ axis. 
In general, two kinds of inversion symmetries exist in lattice systems: 
Bond-centered and site-centered inversion symmetries. 
The dimerization parameter $\delta$ violates the site-centered inversion symmetry, 
while the staggered field $h_s$ breaks the bond-centered one. 
A finite dimerization often occurs in low-dimensional systems with Peierls instability~\cite{Giamarchi03}, 
and a staggered field $h_s$ is known to emerge in a class of quasi one-dimensional (1D) magnets 
with low crystal symmetry~\cite{Dender97,Oshikawa99,Feyerherm00,Umegaki09,Affleck99}. 
In addition, if we consider weakly coupled spin chains, 
a focused spin chain can be approximated by an 1D model subject to a staggered field 
due to the N\'eel order~\cite{Scalapino75,Schulz96,Sato04,Okunishi07,Ishimura80,Coldea10} 
of neighboring chains. 
The model (\ref{eq:chain}) is depicted in Fig.~\ref{fig:models}(a).

The 1D model~(\ref{eq:chain}) is exactly fermionized through 
Jordan-Wigner (JW) transformation~\cite{Giamarchi03,Gogolin04,Tsvelik04} . 
By introducing fermion operators $f_j= e^{-i\pi\sum_{\ell=1}^{j-1}S_j^+S_j^-}S_j^-$ and 
$f_j^\dagger= S_j^+ e^{i\pi\sum_{\ell=1}^{j-1}S_j^+S_j^-}$ with $S^\pm_j= S_j^x\pm iS_j^y$, 
Eq.~\eqref{eq:chain} is mapped to 
\begin{align}
H_{\rm 1D}=\sum_j&\frac{J}{2}(1+(-1)^j\delta)\left(f_{j+1}^\dagger f_j+f_{j}^\dagger f_{j+1}\right)
+(h+(-1)^jh_s)n_j,
\label{eq:JWchain}
\end{align}
where $n_j=f_j^\dagger f_j$ is the number operator on $j$th site. 
The mapped model~\eqref{eq:JWchain} is a 1D tight-binding model of the spinless fermion. 
Its energy bands are shown in Fig.~\ref{fig:models}(b) and (c). 
We have a gapless energy dispersion in the inversion-symmetric case with $\delta=h_s=0$, 
as shown in Fig.~\ref{fig:models}(b), 
while a gap opens and valence and conducting bands appears if we introduce 
an inversion-asymmetric perturbation $\delta\neq 0$ or $h_s\neq 0$.  
When the magnetic field $h$ is small enough, the ground state of the spin chain~(\ref{eq:chain}) 
corresponds to the half-filled state of the fermion model~\eqref{eq:JWchain}. 
We will focus on the half-filled ground state in the rest of this paper. 
Since Eq.~\eqref{eq:JWchain} is viewed as a model of an 1D spinless semiconductor, 
we can take analogy between the photo spin current and usual photo electric current 
in order to study the former of the spin chain. 
%with the help of the established results of usual photo current in semiconductors. 
However, we should note that the fermions $f_j$ and $f_j^\dagger$ represent 
magnetic excitations (not electron) in the spin chain. 
Hereafter, we will call the JW fermions spinons for simplicity.

%%%%%%%%%%%%%%%%%%%%%%%%%%%%
%%%%%%%%%%%%%%%%%%%%%%%%%%%%
\subsection{Antiferromagnetic or Ferrimagnetic Insulators}
\label{sec:AF}
Secondly, we consider the three-dimensional (3D) spin model 
with antiferro or ferrimagnetically ordered ground state. The model is shown in Fig.~\ref{fig:models}(d). 
The Hamiltonian is given by 
\begin{align}
H_{\rm 3D}=&\sum_\textit{\textbf{r}} 
J(1+\delta)\textit{\textbf{S}}_a(\textit{\textbf{r}})\cdot\textit{\textbf{S}}_b(\textit{\textbf{r}})
+J(1-\delta)\textit{\textbf{S}}_a(\textit{\textbf{r}}+\hat x)\cdot\textit{\textbf{S}}_b(\textit{\textbf{r}})
-(D+D_s) \left[S_a^z(\textit{\textbf{r}})\right]^2-(D-D_s)\left[S_b^z(\textit{\textbf{r}})\right]^2\nonumber\\
%%%%%
&-J_\perp\Big[\textit{\textbf{S}}_a(\textit{\textbf{r}})\cdot\textit{\textbf{S}}_a(\textit{\textbf{r}}+\hat y)
+\textit{\textbf{S}}_a(\textit{\textbf{r}})\cdot\textit{\textbf{S}}_a(\textit{\textbf{r}}+\hat z)
+\textit{\textbf{S}}_b(\textit{\textbf{r}})\cdot\textit{\textbf{S}}_b(\textit{\textbf{r}}+\hat y)
+\textit{\textbf{S}}_b(\textit{\textbf{r}})\cdot\textit{\textbf{S}}_b(\textit{\textbf{r}}+\hat z)\Big]\nonumber\\
&-B\Big[g_aS_a^z(\textit{\textbf{r}})+g_bS_b^z(\textit{\textbf{r}})\Big].
\label{eq:model3d}
\end{align}
Here, $\textit{\textbf{S}}_{a(b)}(\textit{\textbf{r}})$ is the spin-$S_{a(b)}$ operator on $\textit{\textbf{r}}$th site 
on $a(b)$ sublattice, where $\textit{\textbf{r}}=(m_x\hat x,m_y\hat y,m_z\hat z)$ with $m_{x,y,z}$ being an integer 
and $\hat a$ ($a=x,y,z$) is the unit vector along the $a$ direction. 
$J$ is the antiferromagnetic exchange coupling constant along the $x$ direction 
and $J_\perp$ is the ferromagnetic constant along $y$ and $z$ directions. 
The parameter $\delta$ shows the dimerization on the $J$ bond. 
We have introduced uniform [staggered] single-ion anisotropy with $D>0$ [$D_s(< D)$] to stabilize 
the antiferromagnetic ordering. 
The final term of Eq.~\eqref{eq:model3d} is the Zeeman interaction induced by a uniform magnetic field $B$, 
and $g_a$ and $g_b$ respectively denote g factors on $a$ and $b$ sublattices.  
The ground state of the model~\eqref{eq:model3d} is N\'eel ordered for $S_a=S_b$ 
and ferrimagnetically ordered for $S_a\neq S_b$, as shown in Fig.~\ref{fig:models}(d). 
These antiferromagnetically ordering patterns break the bond-centered inversion symmetry 
even without a staggered anisotropy $D_s$ or a difference of g factors $g_a-g_b$. 
This is in contrast with the case of the 1D model~\eqref{eq:chain}, 
where a staggered field $h_s$ is necessary to lead to the breakdown of the bond-centered inversion symmetry. 
On the other hand, the dimerization $\delta$ breaks the site-centered inversion symmetry 
of Eq.~\eqref{eq:model3d} like the 1D case.

We can apply the spin-wave theory~\cite{Yoshida96,Mattis06} 
for the antiferromagnetically ordered state of Eq.~\eqref{eq:model3d}. 
Through the following Holstein-Primakoff transformation, 
\begin{align}
S_a^z(\textit{\textbf{r}})=S_a-n_a(\textit{\textbf{r}}),\,\,\,\,\,\,\,\,
S_a^+(\textit{\textbf{r}})=\sqrt{2S_a}\left(1-\frac{n_a(\textit{\textbf{r}})}{2S_a}\right)^{\frac12}
a(\textit{\textbf{r}}),\,\,\,\,\,\,\,\,
S_a^-(\textit{\textbf{r}})=\sqrt{2S_a}a^\dagger(\textit{\textbf{r}})
\left(1-\frac{n_a(\textit{\textbf{r}})}{2S_a}\right)^{\frac12},  \\
S_b^z(\textit{\textbf{r}})=n_b(\textit{\textbf{r}})-S_b,\,\,\,\,\,\,\,\,
S_b^+(\textit{\textbf{r}})=\sqrt{2S_b}b^\dagger(\textit{\textbf{r}})
\left(1-\frac{n_b(\textit{\textbf{r}})}{2S_b}\right)^{\frac12},\,\,\,\,\,\,\,\,
S_b^-(\textit{\textbf{r}})=\sqrt{2S_b}\left(1-\frac{n_b(\textit{\textbf{r}})}{2S_b}\right)^{\frac12}b(\textit{\textbf{r}}),
\end{align}
the spin model~\eqref{eq:model3d} is mapped to a boson model. Here, $a(\textit{\textbf{r}})$ and 
$b(\textit{\textbf{r}})$ are respectively the bosonic annihilation operators at $\textit{\textbf{r}}$th site 
on $a$ and $b$ sublattices. $n_a(\textit{\textbf{r}})=a^\dagger(\textit{\textbf{r}})a(\textit{\textbf{r}})$ and 
$n_b(\textit{\textbf{r}})=b^\dagger(\textit{\textbf{r}})b(\textit{\textbf{r}})$ are the number operators. 
Within the linear spin-wave approximation, the bosonized Hamiltonian is given by 
\begin{align}
H_{\rm 3D}\approx &H_{\rm sw} =\sum_{\textit{\textbf{k}}}
\left(\begin{array}{c}
a_{\textit{\textbf{k}}} \\
b_{-\textit{\textbf{k}}}^\dagger
\end{array}\right)^\dagger
\left(\begin{array}{cc}
h^0_\textit{\textbf{k}}+h^z_\textit{\textbf{k}} & h^x_\textit{\textbf{k}}-{\rm i}h^y_\textit{\textbf{k}} \\
h^x_\textit{\textbf{k}}+{\rm i}h^y_\textit{\textbf{k}} & h^0_\textit{\textbf{k}}-h^z_\textit{\textbf{k}}
\end{array}\right)
\left(\begin{array}{c}
a_{\textit{\textbf{k}}} \\
b_{-\textit{\textbf{k}}}^\dagger
\end{array}\right)+\text{const.}
\label{eq:SW}
\end{align}
Here, $h^{0,x,y,z}_\textit{\textbf{k}}$ are functions of wave vector $\textit{\textbf{k}}$ 
and parameters in Eq.~\eqref{eq:model3d}, and 
we have introduced the Fourier transformation of Holstein-Primakoff bosons as
$a_{\textit{\textbf{k}}}=(1/\sqrt{N})\sum_{\textit{\textbf{r}}} a(\textit{\textbf{r}})
e^{{\rm i}\textit{\textbf{k}}\cdot\textit{\textbf{r}}}$ and 
$b_{\textit{\textbf{k}}}=(1/\sqrt{N})\sum_{\textit{\textbf{r}}} b(\textit{\textbf{r}})
e^{{\rm i}\textit{\textbf{k}}\cdot(\textit{\textbf{r}}+\hat x/2)}$, where $N$ is the total site number of $a$ or $b$ sublattice.
For this bilinear bosonic model, performing the following Bogoliubov transformation,
\begin{align}
a_{\textit{\textbf{k}}}=\cosh\Theta_{\textit{\textbf{k}}}\alpha_{\textit{\textbf{k}}}
+\sinh\Theta_{\textit{\textbf{k}}}\beta^\dagger_{-\textit{\textbf{k}}},
\,\,\,\,\,&
b_{-\textit{\textbf{k}}}^\dagger=\sinh\Theta_{\textit{\textbf{k}}}e^{i\Phi_{\textit{\textbf{k}}}}\alpha_{\textit{\textbf{k}}}
+\cosh\Theta_{\textit{\textbf{k}}}e^{i\Phi_{\textit{\textbf{k}}}}\beta_{-\textit{\textbf{k}}}^\dagger, 
\label{eq:Bogoliubov}
\end{align}
we arrive at the diagonalized Hamiltonian,
\begin{align}
H_{\rm sw} = &\sum_{\textit{\textbf{k}}}E_\alpha(\textit{\textbf{k}})
\alpha_{\textit{\textbf{k}}}^\dagger\alpha_{\textit{\textbf{k}}}
+E_\beta(\textit{\textbf{k}})\beta_{\textit{\textbf{k}}}^\dagger\beta_{\textit{\textbf{k}}}+\text{const.}
\label{eq:SWfinal}
\end{align}
Here, $\alpha_{\textit{\textbf{k}}}$ and $\beta_{\textit{\textbf{k}}}$ are annihilation operators of new bosons (i.e., magnons). 
When $h_+=D(S_a+S_b-1)+D_s(S_a-S_b)-B(g_a-g_b)/2=0$, energy dispersions of two magnons 
$\alpha_{\textit{\textbf{k}}}$ and $\beta_{\textit{\textbf{k}}}$ are degenerate as shown in Fig.~\ref{fig:models}(e). 
For $h_+\neq 0$, the band degeneracy is lifted as in Fig.~\ref{fig:models}(f).

%%%%%%%%%%%%%%%%%%%%%%%%%%%%
%%%%%%%%%%%%%%%%%%%%%%%%%%%%
\subsection{Spin-light couplings}
\label{sec:spin-light}
Let us consider the spin-light coupling to discuss the photo-induced spin current. 
In vacuum, the fundamental interaction between electron spins and electromagnetic fields is the Zeeman coupling. 
However, additional spin-light couplings may emerge in materials through complex interactions 
among light and matter fields.

For the spin chain model~\eqref{eq:JWchain}, 
we consider the following three sorts of spin-light couplings. 
\begin{align}
  H_\text{Z}=&-B(t)\sum_j(\eta-(-1)^j\eta_s)S_j^z.
\label{eq:Zeeman}
\end{align}
\begin{align}
  H_\text{iDM} =& E_y(t)\sum_j (p+(-1)^j p_s)\left(\textit{\textbf{S}}_j\times\textit{\textbf{S}}_{j+1}\right)^z.
\label{eq:iDM}
\end{align}
\begin{align}
H_\text{ms}=E_x(t)\sum_j (A+(-1)^jA_s)(S_j^x S_{j+1}^x+S_j^y S_{j+1}^y).
\label{eq:ms}
\end{align}
Equation~\eqref{eq:Zeeman} represents an AC Zeeman coupling with $B(t)$ being the AC magnetic field 
of an applied electromagnetic wave. Here, we assume that the AC field $B(t)$ is 
parallel to the $S^z$ direction, i.e., longitudinal. 
The remaining two terms of Eqs.~\eqref{eq:iDM} and \eqref{eq:ms} are sorts of magnetoelectric 
couplings~\cite{Tokura14}. 
Equation~\eqref{eq:iDM} is called an AC inverse Dyzaloshinskii-Moriya (iDM) interaction~\cite{Tokura14,Katsura05} and 
it often appears in multiferroic magnets, in which electric polarization is strongly coupled to electron spins. 
The spin chain is located along the $x$ direction and $E_\alpha(t)$ is the $\alpha$ component of the AC electric field 
of an applied wave. 
Equation~\eqref{eq:ms} is called an magnetostriction (MS) type interaction~\cite{Tokura14}, 
where the AC electric field is coupled to 
the local magnetic exchange interaction. A typical origin of Eq.~\eqref{eq:ms} is the spin-phonon coupling.     
Since we focus on inversion-asymmetric spin chains, we have introduced 
both uniform and staggered coupling constants in all the spin-light couplings such as $\eta$ and $\eta_s$.

For the antiferromagnetic model~\eqref{eq:model3d}, we consider a standard AC Zeeman coupling, 
\begin{align}
  H_{\text{Z,3D}}=&-B(t)\sum_{\textit{\textbf{r}}}
\Big[g_aS^z_a(\textit{\textbf{r}})+g_bS^z_b(\textit{\textbf{r}})\Big]
\label{eq:Zeeman2}
\end{align}
We set the AC magnetic field to be parallel to the $S^z$ direction like the 1D case.

We note that the $z$ component of total spin $S_{\rm tot}^z=\sum_j S_j^z$ (or 
$=\sum_{\textit{\textbf{r}}}S^z_a(\textit{\textbf{r}})+S^z_b(\textit{\textbf{r}})$) is conserved in all of our spin-light couplings. 
It means that the angular momentum transfer from photons to spins does not take place in our models. 
Nevertheless, as we will see soon later, an DC spin current is produced through these spin-light couplings.
As one sees from Eqs.~\eqref{eq:Zeeman}-\eqref{eq:Zeeman2}, 
we also note that we focus on simple linearly-polarized electromagnetic waves in this study.

%%%%%%%%%%%%%%%%%%%%%%%%%%%%
%%%%%%%%%%%%%%%%%%%%%%%%%%%%
%%%%%%%%%%%%%%%%%%%%%%%%%%%%
\section{Second-order Response Theory and Results}
\label{sec:result}
In the previous section, we have explained our setup of generating photo-induced spin current. 
This section is devoted to showing our theoretical approach and results. 

%%%%%%%%%%%%%%%%%%%%%%%%%%%%
%%%%%%%%%%%%%%%%%%%%%%%%%%%%
\subsection{Second-order Response}
\label{sec:2nd-order}
%As we mentioned in Sec.~\ref{sec:inversion}, 
The leading term of the DC response to an external AC field emerges 
from the second-order perturbation of the AC field (see Appendix).  
To calculate the DC spin current in our driven magnetic models, 
we apply the second-order response theory~\cite{Kraut79,Belinicher82,Kral00,Sipe00} 
that is a standard extension of the well-known linear-response theory 
(i.e., Kubo formula). It is based on the perturbation expansion of the equation of motion for density matrices 
with respect to an applied AC field. 
Let us represent the AC perturbation term as $H'(t)=F(t){\cal O}(t)$, 
where $F(t)$ is the external AC field and ${\cal O}(t)$ is an operator of the system we consider. 
According to the second-order response theory, the response of a current $I(t)$ is computed as 
\begin{align}
\langle I(t) \rangle_2 = & -\int_0^\infty dt_1dt_2
\Big\langle[{\cal O}(t-t_1-t_2),\,\,[{\cal O}(t-t_1),\,\, I(t)]]\Big\rangle
F(t-t_1-t_2)F(t-t_1),
\label{eq:2nd-order}
\end{align}
where $\langle \cdots\rangle_2$ stands for the second-order response. 
When the Fourier transformation along time $t$ is defined as 
$\langle I(\omega) \rangle_2=\int dt \langle I(t) \rangle_2 e^{-i\omega t}$, 
Eq.~\eqref{eq:2nd-order} is transformed to 
\begin{align}
\langle I(\Omega) \rangle_2 =& \int d\omega \sigma^{(2)}(\Omega;\omega,\Omega-\omega)
F(\omega)F(\Omega-\omega).
\label{eq:2nd_Fourier}
\end{align}
This equation defines the nonlinear conductivity $\sigma^{(2)}(\Omega;\omega,\Omega-\omega)$. 
If we apply this formula to electric current $I(t)$ of an 1D non-interacting electron system, 
the nonlinear DC conductivity ($\Omega=0$) is calculated as 
\begin{align}
\sigma^{(2)}(0;\omega,-\omega)=& \sum_{\alpha,\beta,\gamma}\int\frac{dk}{(2\pi)^2}
\frac{[f_{\alpha}(k)-f_{\beta}(k)]{\cal O}_{\alpha\beta}(k)}
{\omega-\varepsilon_{\beta}(k)+\varepsilon_{\alpha}(k)-i/(2\tau)}
\left[\frac{{\cal O}_{\beta\gamma}(k) I_{\gamma\alpha}(k)}
{\varepsilon_{\alpha}(k)-\varepsilon_{\gamma}(k)-i/(2\tau)}-
\frac{I_{\beta\gamma}(k){\cal O}_{\gamma\alpha}(k)}
{\varepsilon_{\gamma}(k)-\varepsilon_{\beta}(k)-i/(2\tau)}\right],
\label{eq:nonlinear}
\end{align}
where ${\cal O}_{\alpha\beta}(k)=\left\langle \alpha k\right|{\cal O}\left|\beta k\right\rangle$ and 
$I_{\alpha\beta}(k)=\left\langle \alpha k\right| I \left|\beta k\right\rangle$. where 
$|\alpha k\rangle$ is the electron energy eigenstate with wave number $k$ on the $\alpha$ band. 
$\epsilon_\alpha(k)$ is the eigen energy of the state $|\alpha k\rangle$, and 
$f_\alpha(k)=1/(1+e^{\epsilon_\alpha(k)/(k_BT)})$ is the fermion distribution function for $|\alpha k\rangle$. 
We have introduced a phenomenological relaxation time $\tau$ which is important 
to obtain a physically meaningful value of the conductivity $\sigma^{(2)}$. 
Note that $\sigma^{(2)}$ is ill-defined in the case of $1/\tau=0$.

%%%%%%%%%%%%%%%%%%%%%%%%%%%%
%%%%%%%%%%%%%%%%%%%%%%%%%%%%
\subsection{Nonlinear Conductivity of Spin Current}
\label{sec:conduct}
\begin{figure} [ht]
\begin{center}
\begin{tabular}{c} %% tabular useful for creating an array of images 
\includegraphics[height=9cm]{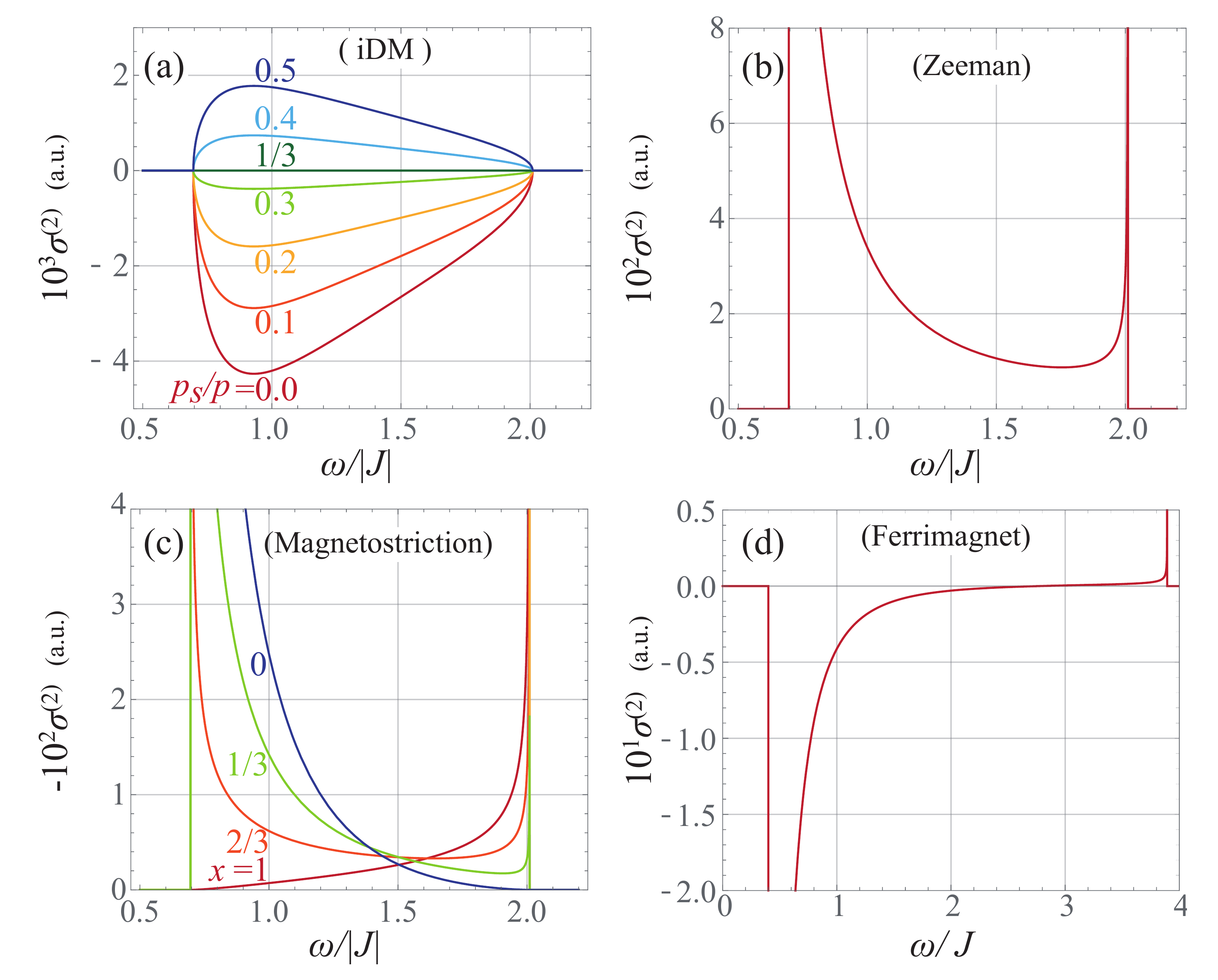}%{mcr3b.eps}
\end{tabular}
\end{center}
\caption{\label{fig:nonlinear} Nonlinear spin-current conductivity at zero temperature 
as a function of the frequency $\omega$ of applied linearly-polarized electromagnetic wave. 
Panels (a), (b) and (c) are the results of the inversion-asymmetric spin-chain model with iDM, Zeeman, and MS type 
spin-light couplings, respectively. 
Parameter $x$ denotes the ratio of uniform and staggered coupling constants: $A=x$ and $A_s=1-x$. 
Panel (d) shows the conductivity of the inversion-asymmetric antiferromagnetically ordered model. 
We take the 1D limit $J_\perp/J\to 0$ and set $g_a\neq g_b$ and $\delta\neq 0$.  
In all the panels, we have taken the ideal limit $1/\tau\to +0$ 
(Note that $1/\tau\to +0$ is not equal to $1/\tau=0$).}
\end{figure}

We can straightforwardly apply the formula~\eqref{eq:nonlinear} to the photo-induced spin current 
in our spin chain model because the Hamiltonian is mapped to a non-interacting 
JW-fermion (spinon) form~\eqref{eq:JWchain}. 
The spin current with $S^z$ polarization may be defined from 
the equation of motion for the $z$ component of total spin. The resultant current is 
\begin{align}
J_{\rm sc}=& \frac{1}{N_{\rm 1D}}\sum_j -J \left[1+(-1)^j \delta\right](S_j^xS_{j+1}^y-S_j^yS_{j+1}^x),
\label{eq:spincurrent}
\end{align}
where $N_{\rm 1D}$ is the total site number. 
Therefore, the real part of the DC nonlinear conductivity is calculated as 
\begin{align}
{\rm Re}[\sigma^{(2)}(0;\omega,-\omega)]=&
\frac1{\pi}{\rm Re}\left\{\sum_{k}\frac{{\cal O}_{+-}(k)J_{-+}(k)[{\cal O}_{--}(k)-{\cal O}_{++}(k)]}
{\omega^2-[\varepsilon_{+}(k)-\varepsilon_{-}(k)-i/(2\tau)]^2}\right\},
\label{eq:spincurrent2}
\end{align}
where $\epsilon_-(k)$ and $\epsilon_+(k)$ are respectively the valence and conduction band dispersions 
[see Fig.~\ref{fig:models}(c)]. $J_{\alpha\beta}(k)$ is the matrix element of the spin current $J_{\rm sc}$, and 
${\cal O}_{\alpha\beta}(k)$ is the matrix element of the operator part of the spin-light coupling 
[see Eqs.~\eqref{eq:Zeeman}-\eqref{eq:ms}]. We note that the imaginary part of the conductivity is zero 
for linearly polarized waves.
%$\cal O}_{\alpha\beta}(k)=\left\langle \alpha k\right|{\cal O}\left|\beta k\right\rangle$
Figure~\ref{fig:nonlinear}(a)-(c) represent the DC nonlinear conductivity of spin current 
in the inversion-asymmetric spin chain model~\cite{Ishizuka19-1} with different spin-light couplings at $k_BT=0$. 
We have set $1/\tau\to +0$ in Fig.~\ref{fig:nonlinear}. 
Three panels (a)-(c) show that a finite nonlinear conductivity occurs for all of the three spin-light couplings 
when the frequency $\omega$ is in the range of the energy difference between valence and conducting bands, 
i.e., when particle-hole pair creation is possible via the photon absorption [see Fig.~\ref{fig:models}(c)]. 
This frequency is usually in GHz or THz regime since the JW-fermion band width is mainly determined 
by the exchange couplings.  
We verify that the conductivity $\sigma^{(2)}$ vanishes unless both the site- and bond-centered symmetries are broken. 
In the case of AC Zeeman interaction or MS one with $A_s\neq 0$ [panels (b) and (c)], 
the conductivity $\sigma^{(2)}$ diverges near the lowest and highest frequencies $\omega$. 
This behavior stems from the divergent nature of the density of state of JW fermions near the band edges. 
We also note that only the staggered coupling constant $\eta_s$ contributes to the photo spin current 
in the AC Zeeman case because the uniform AC Zeeman term conserves $S^z_{\rm tot}$. 
From the frequency dependence of $\sigma^{(2)}$, one can distinguish the kinds of spin-light couplings.

In Ref.~\citenum{Ishizuka19-2}, 
we have extended the formula of Eq.~\eqref{eq:nonlinear} to the bosonic (magnon) systems. 
As a result, we compute the photo-induced spin current of the inversion-asymmetric 
antiferromagnetic or ferrimagnetic insulators~\eqref{eq:model3d}. 
Figure~\ref{fig:nonlinear}(d) shows the conductivity $\sigma^{(2)}$ under the condition of 
$k_BT=0$, $1/\tau\to +0$, and $J_\perp/J\to 0$. 
The AC Zeeman term of Eq.~\eqref{eq:Zeeman2} creates a magnon pair, 
$\alpha_{\textit{\textbf{k}}}$ and $\beta_{\textit{\textbf{k}}}$, and 
therefore the photo-induced spin current can be finite in the frequency range between 
$\omega=E_\alpha(\textit{\textbf{0}})+E_\beta(\textit{\textbf{0}})$ and 
$E_\alpha(\pi,\pi,\pi)+E_\beta(\pi,\pi,\pi)$. 
We verify that a finite conductivity appears only in the presence of a finite dimerization $\delta$. 
Namely, as expected, the inversion symmetry breaking is the necessary condition 
for the photo-induced magnon spin current. 
We also find that a difference $g_a-g_b$ in the AC Zeeman coupling is necessary 
to lead to a finite conductivity $\sigma^{(2)}$.  
The divergent behavior of $\sigma^{(2)}$ near the lowest and highest frequencies is attributed to 
the magnon large density of state near the band edges like the spinon case.  

In the end of this section, we discuss the relaxation-time ($\tau$) dependence of 
the spin-current conductivity. As we mentioned, Fig.~\ref{fig:nonlinear} depicts the conductivity 
at the ideal limit $1/\tau\to +0$, where quasi particles (spinons or magnons) have a sufficiently long life time. 
We numerically show that the conductivity is quite insensitive against the change of $\tau$. 
Namely, the leading term of the nonlinear conductivity is independent of the relaxation time, 
\begin{align}
{\rm Re}[\sigma^{(2)}(0;\omega,-\omega)]\,\,\,\sim&\,\,\,\text{const.}+\tau\text{-dependent\,\,\,terms}.
\label{eq:shiftcurrent}
\end{align}
This is in sharp contrast with the usual case of injection current, 
in which the leading term of conductivity is proportional to $\tau$.  
The current satisfying the condition~\eqref{eq:shiftcurrent} is called shift current~\cite{Kraut79,Belinicher82,Kral00,Sipe00}. 
Therefore, the predicted spin currents are all shift current type, and 
it means that the photo-induced spin current is robust against impurity scattering.
We emphasize that the calculated magnon shift current is the first example of 
a shift current carried by bosonic quasi particles to the best of our knowledge. 
  
\begin{figure} [ht]
\begin{center}
\begin{tabular}{c} %% tabular useful for creating an array of images 
\includegraphics[height=5cm]{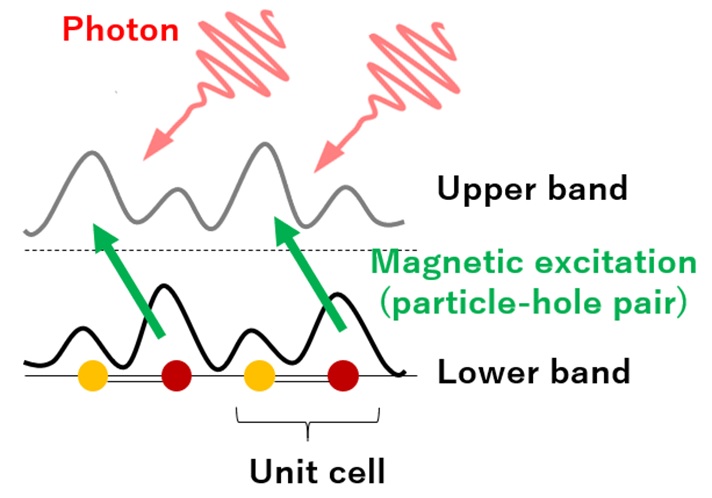}%{mcr3b.eps}
\end{tabular}
\end{center}
\caption{\label{fig:shift} Intuitive real-space picture of the shift spin current 
in our inversion-asymmetric spin chain. Due to the inversion asymmetry, the center of mass of JW fermions 
in both conducting (upper) and valence (lower) bands deviates from the center of unit cell. 
}
\end{figure}

Finally, we explain the intuitive picture of shift current, focusing on the spin chain model~\eqref{eq:chain}. 
Due to the noncentrosymmetry, the shape of Bloch wave functions of JW fermion on conducting and valence bands 
are asymmetric in the unit cell, as shown in Fig.~\ref{fig:shift}. 
If we apply THz or GHz wave with a suitable frequency, particle-hole pairs are created 
as in Figs. \ref{fig:models}(c) and \ref{fig:shift}. This accompanies a ``shift'' of center of mass of JW fermions 
due to the asymmetric shape of the Bloch functions. Therefore, this photo transition makes 
the carrier of spin current move along a certain direction without varying the JW-fermion number, 
i.e., with conserving $S_{\rm tot}^z$. This mechanism is the origin of the word ``shift''.

%%%%%%%%%%%%%%%%%%%%%%%%%%%%
%%%%%%%%%%%%%%%%%%%%%%%%%%%%
%%%%%%%%%%%%%%%%%%%%%%%%%%%%
\section{Experimental Methods, Materials, and Laser Intensity}
\label{sec:observability}
In this section, we discuss our spin-current rectification from the experimental viewpoint. 
Below, we discuss the experimental ways of detecting the photo spin current, 
candidate materials for the rectification, and the required strength of applied THz/GHz waves. 

\begin{figure} [ht]
\begin{center}
\begin{tabular}{c} %% tabular useful for creating an array of images 
\includegraphics[height=4cm]{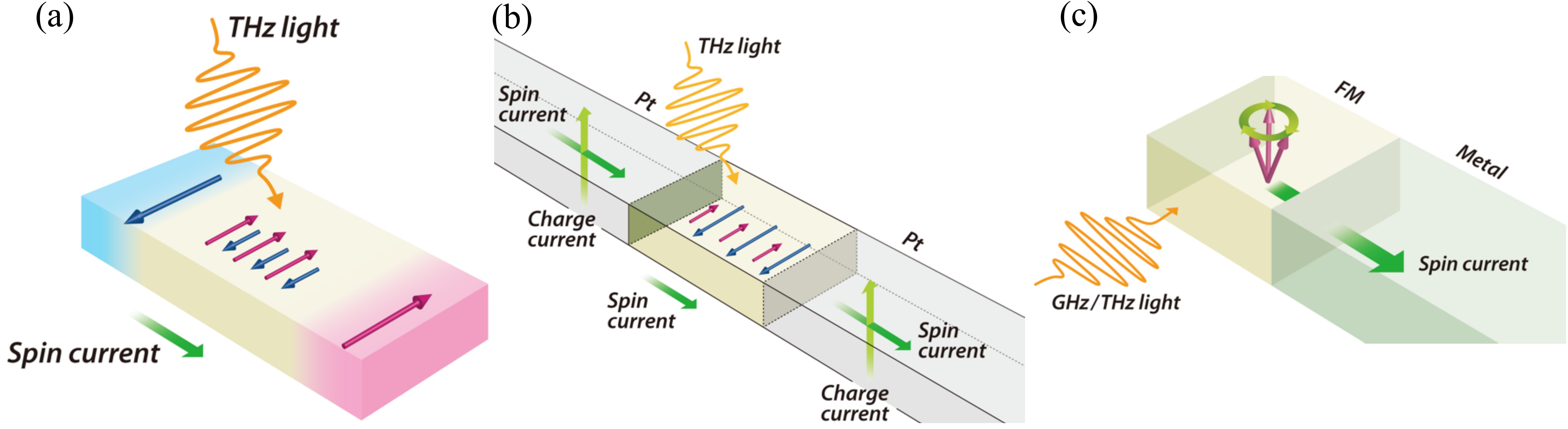}%{mcr3b.eps}
\end{tabular}
\end{center}
\caption{\label{fig:detection} 
Possible experimental methods (a) and (b) of detecting photo-induced spin current of our proposal. 
As a comparison, a standard way of detecting spin current of spin pumping is depicted in the panel (c). }
\end{figure}

First, we consider experimental ways~\cite{Ishizuka19-1,Ishizuka19-2} 
of observing the clear signature of photovoltaic spin current. 
The simplest way of the detection is an all-optical method described in Fig.~\ref{fig:detection}(a). 
The setup is merely to apply a THz/GHz wave to a target magnetic material. 
After the continuous irradiation, a non-equilibrium steady state with a spin current is realized 
and a spin polarization accumulates around one and another edges, as shown in Fig.~\ref{fig:detection}(a). 
The accumulated magnetization can be detected through inverse Faraday or Kerr effect~\cite{Kirilyuk10}. 
Different signs of magnetizations at two edges become a clear evidence of our spin-current rectification. 
   
The second way is based on inverse spin Hall effect~\cite{Saitoh06,Valenzuela06,Kimura07} 
in an attached metal, and it has been often used in the field of spintronics~\cite{Maekawa12}. 
The setup is shown in Fig.~\ref{fig:detection}(b), in which we attach two heavy metals such as Pt and W 
to the target magnet. Due to the directionality of the photovoltaic spin current, 
the spin current in the magnet near one interface is injected into a metal, 
while the spin current near another interface is absorbed from another metal to the magnet. 
A part of the spin current in metals changes into an electric current via inverse spin Hall effect, 
and then we observe an electric voltage perpendicular to the spin-current direction. 
The signs of electric voltages in two attached metals give us a clear evidence of the spin-current rectification.

Next, we mention candidate materials for the photovoltaic spin current.   
Our theoretical analysis indicates that the photovoltaic spin current can be created 
in a broad class of inversion-asymmetric magnets. For instance, 
we can find the following inversion-asymmetric magnets: 
a magnetoelectric material $\rm Cr_2O_3$~\cite{McGuire56}, 
ferrimagnetic diamond chains~\cite{Okamoto03,Shores05,Vilminot06},
multiferroic materials~\cite{Murakawa10,Viirok19}, and a polar ladder magnet $\rm BaFe_2Se_3$~\cite{Aoyama19}.
In addition, meta magnetic materials without inversion symmetry would also be a candidate.
As we showed in Sec.~\ref{sec:conduct}, a large density of state in quasi-1D materials has a high potential 
to enhance the value of the nonlinear spin-current conductivity. 
Therefore, low-dimensional noncentrosymmetric magnets are expected to be a nice candidate 
for a large photovoltaic spin current.

Finally, we shortly discuss the required intensity of applied THz/GHz waves for the realization of 
the spin-current rectification. We have estimated it within a semi-quantitative level 
in Refs.~\citenum{Ishizuka19-1,Ishizuka19-2}. 
According to our estimate, an AC electric fields of $10^{2-5}$ V/cm and $10^{4-6}$ V/cm are sufficient 
to produce an observable spin current in the asymmetric spin chains in Eq.~\eqref{eq:chain} and 
the antiferromagnetically ordered insulators in Eq.~\eqref{eq:model3d}, respectively. 
One can reach these intensities by using currently-available THz laser techniques.

%%%%%%%%%%%%%%%%%%%%%%%%%%%%
%%%%%%%%%%%%%%%%%%%%%%%%%%%%
%%%%%%%%%%%%%%%%%%%%%%%%%%%%
\section{Summary and Discussions}
\label{sec:summary}
We explained our recent theoretical proposal~\cite{Ishizuka19-1,Ishizuka19-2} of the spin-current rectification 
in inversion-asymmetric magnetic insulators. 
We show that DC spin current can be produced with a linearly polarized THz or GHz wave in 
inversion-asymmetric spin chains in Eq.~\eqref{eq:chain} with spinon (fermionic) excitations 
and antiferromagnetically ordered insulators in Eq.~\eqref{eq:model3d} with magnon (bosonic) excitations. 
Our result indicates that the spin-current rectification is possible in a wide class of noncentrosymmetric magnets. 
We show that the photovoltaic spin current is of a shift current and stable against impurity scattering.

Thanks to the long history of magnetism, various magnetic materials have been discovered and synthesized. 
However, unfortunately, only a few kinds of ordered magnets such as ferro and antiferomagnets 
have been investigated in the field of photo-induced phenomena. 
Photo-driven phenomena in other various magnetic materials have a high potential to open a new field 
of magneto-optics and photonics.

Finally, we compare our proposal of the spin-current rectification 
with the well-established spin pumping effect~\cite{Kajiwara10,Heinrich11}. The latter effect is a representative of 
producing a DC spin current with external electromagnetic waves. 

\begin{figure} [ht]
\begin{center}
\begin{tabular}{c} %% tabular useful for creating an array of images 
\includegraphics[height=5cm]{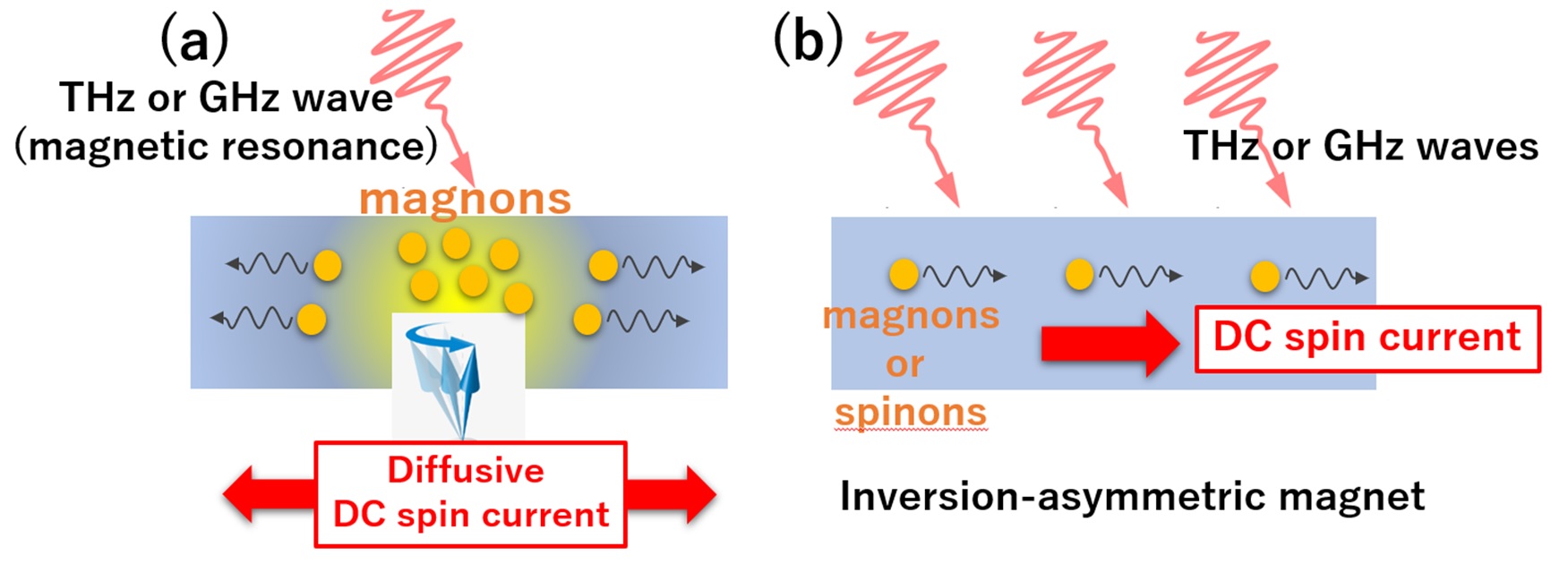}%{mcr3b.eps}
\end{tabular}
\end{center}
\caption{\label{fig:spinpump} Schematic setups of 
(a) spin pumping and (b) spin-current rectification (our proposal). 
The setup of spin pumping is basically the same as the standard magnetic resonance. 
Due to the GHz or THz wave-driven magnetic resonance, 
a plenty of magnons are created and then diffusively spread. 
On the other hand, in the case of the rectification (b), a spin current is created 
by photo-induced magnetic excitations and its flow direction is uniquely determined 
by the inversion asymmetry of the crystal.  
}
\end{figure}

A typical setup of spin pumping is described in Fig.~\ref{fig:detection}(c) and 
is basically the same as a standard magnetic resonance in magnets, as shown in Fig.~\ref{fig:spinpump}(a). 
If we apply GHz or THz wave with the resonant frequency to an ordered magnet, 
the magnetic resonance occurs and many magnons are created. 
This is an angular momentum transfer from photons to magnons, namely, the value of $S_{\rm tot}^z$ is changed. 
Created magnons diffusively spread in all directions as in Fig.~\ref{fig:spinpump}(a), and 
it means the production of a diffusive spin current. For the detection of the diffusive spin current, 
the attachment of a heavy metal has been often used like our proposed method in Fig.~\ref{fig:detection}(b). 
However, the spin-pumped spin current is always injected into the metal 
irrespective of the direction of the interface between the magnet and the metal. 

On the other hand, as we discussed in the previous section, 
the photovoltaic spin current propagates along a certain direction, 
which determines from the inversion-symmetry breaking of the crystal, as shown in Fig.~\ref{fig:spinpump}(b). 
This directionality nature is essentially different from diffusion dynamics of spin pumping. 
Therefore, the attachment of a heavy metal is not always necessary 
to detect photo-induced spin current, in principle. In addition, as we already mentioned, 
the photo spin current does not accompany the change of $S_{\rm tot}^z$ and is shift current type. 
Moreover, the AC-field polarization dependence would be useful to distinguish spin current rectification 
and spin pumping. 

From these arguments, one sees that the predicted spin current rectification is 
essentially different from the established spin pumping phenomenon. 
There is a possibility that spin currents already observed in spin pumping experiments 
include the photovoltaic spin current.

%%%%%%%%%%%%%%%%%%%%%%%%%%%%%%%%%%%%%%%%%
%%%%%%%%%%%%%%%%%%%%%%%%%%%%%%%%%%%%%%%%%
%\bibliography{main_v0.4_revtex_jpg}

%\begin{thebibliography}{12}
%\end{thebibliography}

\end{document}